# Testing existence of antigravity

Dragan Slavkov Hajdukovic
CERN, Geneva, Switzerland
and Cetinje, Montenegro
dragan.hajdukovic@cern.ch

After a brief review of arguments in favor of antigravity (as gravitational repulsion between matter and antimatter) we present a simple idea for an experimental test using antiprotons. Different experimental realizations of the same basic idea are considered

## *1. Introduction*

The weak equivalence principle (WEP) is the oldest and the most trusted principle of contemporary physics. Its roots go back to the time of Galileo and his discovery of "universality of the free fall" on Earth. A deeper understanding was achieved by Newton who explained "universality of the free fall" as a consequence of the equivalence of the inertial mass $m_I$ and the gravitational mass $m_G$. Unlike many other principles, which lose importance with time, the WEP has even increased its importance and is presently the cornerstone of Einstein's General Relativity and of modern Cosmology. It is amusing that physicists use the word "weak" for a principle having such a long and successful life and such great names associated with it (Galileo, Newton, and Einstein). It should rather be called "the principle of giants".

Speculations concerning possible violation of the WEP may be divided into two groups. The first (and older) group consists of a number of different theoretical scenarios for minimal violation of the WEP. The universality of gravitational attraction is not being questioned (so, there is no room for antigravity), but in some cases gravitational and inertial mass may slightly differ (See [1] and References therein). In the present paper we are *exclusively* interested in the second group of speculations, (which has appeared in the last decade of the 20[th] century) predicting gravitational repulsion between matter and antimatter, i.e. antigravity as the most dramatic violation of the WEP.

There are three *new* lines of argument in favour of antigravity.

1. The intriguing possibility that CP violation observed in the neutral Kaon decay may be explained by the hypothesis of a repulsive effect of the earth's gravitational field on antiparticles [2].

2. The Isodual theory of antimatter [3].

3. The growing evidence (since 1998) that the expansion of the universe is accelerating [4] rather than decelerating. Antigravity may be a possible explanation of this accelerated expansion of the universe [5].

Just to add a little bit of humor to the subject, I will point out one more argument. Looking at all the natural beauties and wonders on our planet and in the universe, we must conclude that God was a child at the time of Creation – because only a child can have such an imagination. And, if at that time God was a child, antigravity must exist!

Only 10 years ago, the scientific community was nearly unanimously considering antigravity as a bizarre idea. But today, in view of new observations and arguments, experimental tests to prove or disprove the existence of antigravity are *a must* for both General Relativity and Cosmology.

The first and up to now, last attempt to measure the gravitational acceleration of antiproton was PS-200 experiment at CERN. *In the absence of better ideas*, PS-200 was based on the use of TOF (time-of-flight) technique [6]. After the unsuccessful conclusion of PS-200, general opinion is that the study of the gravitational proprieties of antimatter must be postponed until the production and spectroscopy of antihydrogen, has been fully mastered, making it possible to test the WEP with a high degree of precision [7]. But why wait for more than a decade? We believe that thanks to the ideas proposed in the following sections, it shall become possible to find out if antigravity exists within the next three years.

## *2. The basic idea for experimental testing of antigravity*

In order to grasp the idea, let us start from the equation of motion of a charge q in an electromagnetic field

$$\frac{d}{dt}\vec{p} = q\vec{E} + q\vec{v} \times \vec{B} \qquad (1)$$

The solution of this equation for a particle with non-relativistic velocity $\vec{v}$, in crossed, constant and uniform electric ($\vec{E}$) and magnetic field ($\vec{B}$) may be found in any good book of electrodynamics (see for example Ref. [8]). The crucial point for further analysis is that in the direction perpendicular to the common plane of fields $\vec{E}$ and $\vec{B}$, a particle moves with a velocity which is a periodic function of time with an average value, drift velocity

$$\vec{v}_{drift} = \frac{\vec{E} \times \vec{B}}{B^2} \qquad (2)$$

In the particular case $\vec{E}=0$, $\vec{v}_{drift}=0$ and the orbit of the particle is a helix, with its axis parallel to $\vec{B}$ (in fact the well known *cyclotron* motion).

Let us consider the case $\vec{E}=0$, but instead there is a constant and uniform gravitational field. That instead of $q\vec{E}$ there is a term $m_I \vec{a}$ in the equation of the motion, $\vec{a}$ being the gravitational acceleration. So, gravitation is equivalent to the existence of an electric field $\vec{E} = m_I \vec{a}/q$ in Eq. (1). The corresponding expression for drift velocity caused by gravitation is

$$\vec{v}_{drift} = \frac{m_I}{q} \frac{\vec{a} \times \vec{B}}{B^2} \qquad (3)$$

In principle, Eq. (3) can serve as basis for determination of the gravitational acceleration of antiprotons. In particular, Eq. (3) is suitable for searching antigravity. In fact, if there is antigravity, both q and $\vec{a}$ change the sign when a particle is replaced by an antiparticle. So, if there is antigravity, both proton and antiproton drift in *the same direction*. If there is no antigravity, they drift in opposite directions.

Let us note that in the case of the inhomogeneous magnetic field, there is magnetic gradient force $\vec{F} = grad(\mu B)$, where $\mu$ (the magnetic moment associated with a cyclotron orbit) is a constant of the motion [9]. So, the corresponding drift velocity is

$$\vec{v}_{drift} = \frac{\mu}{q} \frac{gradB \times \vec{B}}{B^2} \qquad (4)$$

So, in the general case, there are drift velocities (2), (3) and (4) caused respectively by an electric field, a gravitational field and a magnetic field gradient. Of course, from the point of view of the gravitational experiment, non-gravitational drifts (2) and (4) are just unwanted complication.

Let's consider the *ideal experimental configuration.* It is a horizontal, *sufficiently long*, metallic, cylindrical, drift tube, with constant and uniform magnetic field $\vec{B}$ (the only non-gravitational field!) collinear with the axis of the cylinder. Under these ideal experimental conditions, the particle motion is superposition of only three motions: cyclotron motion (in which a charged particle is bound to a magnetic field line); drift motion caused by gravity (or antigravity) and motion with constant velocity $\vec{v}$ collinear to the axis of the cylinder. However, drift velocity (3) is extremely small (in a field B=1T it is about 0.1 micrometer/sec, i.e. ten thousand seconds are needed for drift of one millimeter!). Consequently, a sufficiently long drift tube can't be realized in practice. So, instead of a very long tube, it is necessary to confine the movement of the particle in a short cylindrical tube. But this confining of the particles demands additional non-gravitational fields (electric field or inhomogeneous magnetic field). Additional fields therefore assure, crucial, axial confinement of the particles, but there is high price to pay for it: Instead of a purely gravitational drift, there is a superposition of gravitational drift (3) and non-gravitational drifts (2) and/or (4), caused by the confining fields. Pity! The beauty and simplicity of the direct use of formula (3) are consequently shattered.

In short, the key non-trivial idea is to study gravitational properties of antiprotons, measuring effects of the gravitational drift on particles confined in a trap (or a quasi-trap). This idea was firstly suggested seventeen years ago [10], but at that time (1989) antigravity was not considered a credible idea [11] and the interest for such an experiment was very low.

## 3. Different possibilities for realization of the basic idea

### 3.1 General considerations

Different choices of confining electromagnetic fields lead to different realizations of the same basic idea. *The open question is: which is the most appropriate choice of the confining fields* for the purpose of the gravitational experiment, i.e. how to reduce to a minimum, the inevitable perturbations of the tiny effect of the gravitational drift?

One possible choice is to use confining fields with cylindrical symmetry, so that symmetry is violated only by gravitation. In general, if $\rho, \varphi, z$ are cylindrical coordinates, cylindrical symmetry means, that potential $V$ and the intensity of magnetic field $B$ do not depend on $\varphi$. So, $V = V(\rho, z)$ and $B = B(\rho, z)$. Consequently, electric field ($\vec{E}$) and the magnetic field gradient ($gradB$) have only axial and radial components. Thus, *in absence of gravity*, non-gravitational drift velocities (2) and (4) are always tangential to a circle perpendicular to the geometrical axis of the cylinder and having centre on that axis. The result is a circular precession (known as magnetron motion) around the symmetry axis.

Effects of gravity may be literally switched on and off, putting the drift tube respectively in horizontal and vertical position. When gravity is "switched on" (horizontal drift tube), magnetron motion may be slightly perturbed in two different ways:

Firstly, the resultant force $\vec{F}$ (the vector sum of the gravitational force $m_I \vec{a}$, and radial non-gravitational force $\vec{F}_\rho$) is not radial for a circle with centre on the geometrical axis, but for a circle with the centre shifted *above* the geometrical axis in the case of gravity, and *below* the axis in the case of antigravity. The new centre is the point at which the resultant force $\vec{F}$ vanishes. The corresponding distance $\Delta$ between axis and the new centre is determined by

$$\Delta = \frac{m_I a}{F_\rho / \rho} \qquad (5)$$

So, direction of the shift $\Delta$ contains information if there is antigravity or not.

Magnetron motion (modified by gravity or antigravity) starts with injection of the particle. Figure 1 provides a schematic view of the modified magnetron motion of a particle injected along the axis of the cylinder. The black and red circles correspond respectively to gravity and antigravity. Full lines with an arrow represent corresponding position after some time (in fact a few hours). So, if after some time the particle is detected below the horizontal plane passing through the axis, it proves that there is antigravity. If there is no antigravity, but only violation of the WEP, both circles will have the centre above the axis but with different radii.

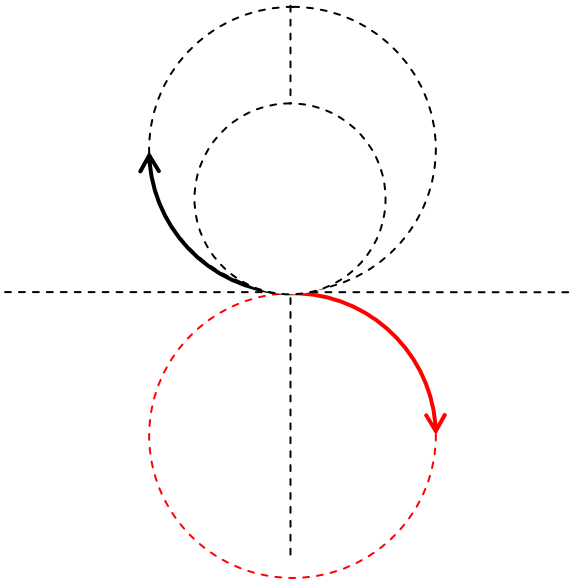

Fig. 1: Schematic view of the magnetron motion modified by gravity or antigravity.

For a particle injected along the axis of the cylinder, black and red circles correspond respectively to gravity and antigravity. For simplicity only the cross section of the cylinder is shown, because axial motion is not essential for the understanding of the principle: if after some time a particle is detected below the horizontal plane passing through the axis, there is evidence of antigravity. A simple violation of the WEP corresponds to two circles with centre above the axis but with different radii.

A second effect, not pertinent to antigravity, is the change of the linear velocity of circular precession (that may be determined from the arc length distance travelled during a known period of time).

## 3.2 Standard electromagnetic traps for charged and neutral particles

Electromagnetic traps for charged and neutral particles, based on the use of electric or magnetic multipole fields, have a few decades of very successful applications in experimental physic [12]. It would be ideal if one of these routinely used traps can serve as realisation of our basic idea. The assessment of existing traps shows that only two of them, the Penning trap and the magnetic bottle, may be appropriate. In a cylindrical Penning trap confinement is achieved with a constant and homogenous magnetic field collinear to the axis of the cylinder, and a quadrupole electrostatic field giving a parabolic potential near the centre of the trap. In a magnetic bottle only an inhomogeneous magnetic field is used.

Our basic idea was studied in detail for both a Penning trap [13] and a magnetic bottle [14]. The comparison between these analyses favours the magnetic bottle. In fact the gravitational effect is completely negligible at typical experimental conditions for Penning traps. The "Gravitational Penning trap" must be much larger and the applied voltage six order lower than in a standard trap. In a magnetic bottle, gravitational and magnetic gradient forces are comparable at typical experimental conditions and consequently a standard trap may be used.

Results for the Penning trap and the magnetic bottle are a certain encouragement for the basic idea. But it is clear that further research for more appropriate confining fields is needed. A common lesson from both analyses is that confining non-gravitational forces mustn't be more than one order of magnitude stronger than the gravitational force (which is about $10^{-7} eV/m$). It is a great challenge to find the suitable trapping configuration under the restriction of such week confining fields.

## 3.3 Search for the most appropriate confinement

A purely magnetic confinement is possible only with an inhomogeneous magnetic field, a principle exploited for the construction of magnetic bottles. Already now, the magnetic gradient force on a particle may be reduced to a value only six times larger than the gravitational force. So, there is no room for the essential improvements of the use of the magnetic confinement as explained in Ref. [14]. Consequently, in this section we will not anymore discuss purely magnetic confinement.

Our basic apparatus is a metallic, cylindrical, maximum two meters long drift tube, with constant and uniform magnetic field $\vec{B}$ collinear with the axis of the cylinder. The field $\vec{B}$ assures both radial confinement of particle and existence of gravitational drift described by Equation (3). It is necessary to add an electric field $\vec{E}$ in order to assure axial confinement (i.e. oscillations of particle between ends of the cylinder). Such fields may be produced in many interesting ways. For instance:
1. The potential of a circular charged ring with uniform linear charge density, placed perpendicularly to the axis of the cylinder, centered on the midpoint of the cylinder. Alternatively, two circular charged rings might be used, placed on opposite sides, but at the same distance of the center of the cylinder.

2. The potential of two end caps of the cylinder with uniform surface charge density. In the limit where the distance between end caps is small compared with their radius, the electric field between them is very similar to the field between two infinite planes with the same charge density. The particle may reflect from the end cap only when it is extremely close to it.
3. A two-dimensional electric field (the two-dimensional quadrupole) as used in Paul traps [12]. If gravity is perpendicular to the plane of a two-dimensional electric field, a net effect of drift velocity (3) may be measured.
4. It may be a bizarre idea, but if the ends caps of the cylinder are close enough the field between them may be managed to be the field of a parallel-plate capacitor (i.e. constant, homogenous and parallel to the axis of the cylinder). If there is periodic change of the voltage of the end caps there is a tiny trapping possibility in a configuration of fields close to the ideal experimental conditions discussed in the second section.

All these and some other possibilities deserve detailed study but here we are more interested in what we call "*minimal confining configuration*".

### 3.3.1 Minimal confining configuration

As it may be concluded by the general formula (5), and in particular from the study of the Penning trap, the quotient $m_1 a / F_\rho$ mustn't be too small. In other words radial electrical force must be very small (close to the value of the gravitational force: $1.6 \times 10^{-26} N$ i.e. $10^{-7} eV/m$). So, the question is: what is the weakest possible confining electrical field? Since there are no electrical fields without charged particles, the question may be reformulated as: *what is the minimal number of charged particles needed to produce a confining electric field*? Just one particle may produce a confining electric field, but in order to preserve cylindrical symmetry, a minimum of two particles must be used. So, as we will demonstrate, the "minimal confining configuration", consists of two charged particles, producing an electric field for the confinement of a third particle (an antiproton in our case).

In order to grasp the idea let's imagine that two antiprotons are somehow "fixed" on the axis of the cylinder at the distances $z = L$ and $z = -L$. It is evident, that the electric field produced by these two antiprotons, together with magnetic field $\vec{B}$, can assure confinement of a third antiproton.

One antiproton may be "fixed" on the axis (i.e. forced to oscillate with very small amplitude) by using an appropriate circular charged ring with uniform linear charge density. This ring must be hidden behind a metal wall, in such a way that its field is shielded from the area where the antiproton used for the gravitational measurement, is moving. So, the antiproton testing antigravity moves only in the electric field of the "fixed" antiprotons. Fig. 2 shows a possible experimental realization. The cylinder is divided into three parts. The central part is partially separated from the other two by metal walls, having in the centre a circular hole. Both non-central parts of the cylinder has a charged ring for "fixing" one antiproton on the axis near the entrance of the central part. The circular metal wall, separating central part of the cylinder, serves to hide the ring.

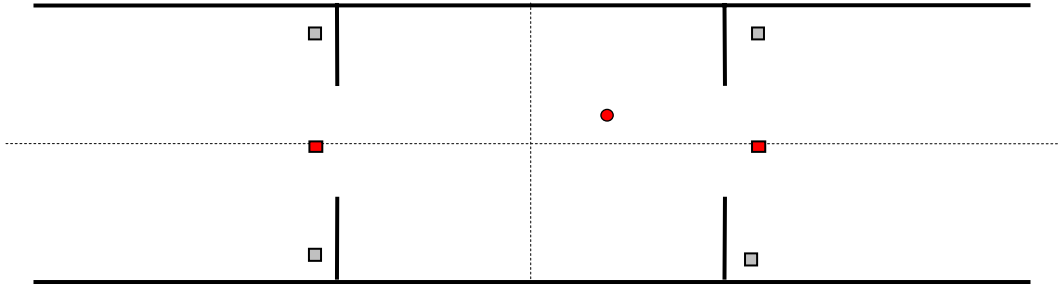

Figure 1: Schematic view of the experimental configuration

Antiproton for testing antigravity (●) moves in a uniform and homogenous magnetic field collinear with the axis of the metallic cylinder, and within the electric field of two antiprotons (■), that are "fixed" on the axis by using the corresponding circular charged ring (■) with uniform linear charge density. Rings are hidden behind metal walls, so that their electric field is shielded from the central part of the cylinder where the antiproton moves.

Without detailed studies (which are nevertheless under way), it is impossible to say if this is a feasible configuration. Just as an indication that it may be feasible, let's estimate the value of $\Delta$ given by formula (5):

Let the origin and the $z$ axis, of the system of the cylindrical coordinates $(\rho,\varphi,z)$ coincide respectively with the centre and the axis of the cylinder, and let two antiprotons "fixed" on the $z$ axes at $z = L$ and $z = -L$. These two "fixed" antiprotons produce a potential, with cylindrical symmetry

$$V(\rho,z) = -\frac{ne}{4\pi\varepsilon_0}\left(\frac{1}{r_1} + \frac{1}{r_2}\right) \quad (6)$$

where $r_1$ and $r_2$ are given by

$$r_1 = \sqrt{(L+z)^2 + \rho^2}$$
$$r_2 = \sqrt{(L-z)^2 + \rho^2} \quad (7)$$

In fact we have written a more general potential with a small number $n$ of "fixed" antiprotons, so that our "minimal confining configuration" corresponds to $n = 1$.

The potential (9) determines radial and axial component of the electric field ($\vec{E}_\rho$ and $\vec{E}_z$) and corresponding forces. In particular, for small values of $\rho$, the radial force is

$$\vec{F}_\rho = \frac{ne^2}{4\pi\varepsilon_0}\left\{\frac{1}{(L+z)^3} + \frac{1}{(L-z)^3}\right\}\vec{\rho} \quad (8)$$

and consequently

$$\Delta = m_I a \Big/ \frac{ne^2}{4\pi\varepsilon_0}\left\{\frac{1}{(L+z)^3} + \frac{1}{(L-z)^3}\right\} \quad (9)$$

In SI units, the numerical value of $e^2/4\pi\varepsilon_0$ is $2.3\times10^{-28}$ Nm$^2$ and $m_I a = 1.6\times10^{26}$ N. By comparing these numbers, it is clear that the achievable value of $F_\rho$ (and consequently of $\Delta$), is much better than in the case using a magnetic bottle (which, by itself, is better than the Penning trap). For instance, $L = 0.25m$ and $z = 0.2m$ give $\Delta \approx 1.8$ cm (with further growth of $\Delta$ for smaller values of $z$).

Even if the final feasibility study for this configuration is negative, it deserves attention because it is the first attempt to attack the problem from opposite side. In fact, electric multipole fields in traditional traps are generated by shaped electrodes extended in space, demanding in principle a distribution of charges that can be considered as continuous. So, the inherent limitation of these configurations is that the number of charged particles producing a confining field must be big, causing trouble with an electric field too strong compared with gravitation. In our "minimal confining configuration", the confining field is not generated by electrodes, but by a low number of charged particles.

### 3.3.2 Concerning realizations without cylindrical symmetry

Configurations, in which the confining electric field has cylindrical symmetry, must be based on formula (5). A confining electric field without cylindrical symmetry may have the advantage of direct use of relation (3) for drift velocity $\vec{v}_{drift}$.

Let the origin and the $z$ axis, of the system of the Cartesian coordinates $(x, y, z)$ coincide respectively with the centre and the axis of a horizontal cylinder and let the $y$ axis be vertical and oriented towards Earth. If a small point-like charge is somehow "fixed" *on the x axis* at a distance $L$ from the origin, it is a simple confining field without cylindrical symmetry, *but is symmetrical in respect to the horizontal plane determined by both the axis x and z*. Potential $V$ and the corresponding electrical forces $\vec{F}_x, \vec{F}_y, \vec{F}_z$ are

$$V = \frac{Q}{4\pi\varepsilon_0}\{(L+x)^2 + y^2 + z^2\}^{-\frac{1}{2}} \tag{10}$$

$$\vec{F}_x = -\frac{eQ}{4\pi\varepsilon_0}\{(L+x)^2 + y^2 + z^2\}^{-\frac{3}{2}}(a+x)\vec{i} \tag{11}$$

$$\vec{F}_y = -\frac{eQ}{4\pi\varepsilon_0}\{(L+x)^2 + y^2 + z^2\}^{-\frac{3}{2}} y\vec{j} \tag{12}$$

$$\vec{F}_z = -\frac{eQ}{4\pi\varepsilon_0}\{(L+x)^2 + y^2 + z^2\}^{-\frac{3}{2}} z\vec{k} \tag{13}$$

It is evident that the tendency of forces $\vec{F}_y$ and $\vec{F}_z$ is to produce oscillations respectively along $y$ and the $z$ axis, while $\vec{F}_x$ causes drift velocity (2) along the $y$ axis. So, gravitational drift velocity (3), which is in the $x$ direction, causes a drift easily distinguishable from the electric drift. It is interesting possibility.

## *4. Comments*

The purpose of this paper is to present the basic idea and to sketch its most promising realizations. The complete feasibility study is under way and if it is positive, we shall request an approval and beam time from the SPSC Committee at CERN by the end of 2006. We estimate that only two beam periods (summer 2007 and summer 2008) are sufficient to conclude the eventual experiment. So, under ideal circumstances, 2008 may be the year of antigravity.

The proposed strategy clearly separates research on antigravity from research on minimal violation of the WEP. So, the first step is to research antigravity and only if antigravity doesn't exist, it makes sense to plan much more complex experiments searching for tiny violation of the WEP. As already pointed out, high precision test on the WEP must be postponed until the production and spectroscopy of the antihydrogen has been fully mastered. However, if the scientific community shows "political" will to do it, an accuracy of a few parts in $10^5$ may be achieved with alternative methods. For instance a drift tube in the form of cycloid may be used (See Reference [15] and Appendix A), with a magnetic field assuring radial confinement and a *gravitational field assuring axial confinement and forcing particle to oscillate along a cycloid.*

Of course, our cylindrical drift tube is just the last missing element in the apparatus for the study of antigravity. The whole apparatus consists of storage, cooling and transfer traps assuring injection of antiprotons (one by one) into the final cylindrical trap, which we have considered in this paper. We did not discuss the other needed parts of apparatus, simply because they are already well developed and used in a series of successful experiments with antiprotons at CERN.

## *Appendix A:*

## *Searching violations of the WEP (in case there is no antigravity)*

The experiments based on the effects of the gravitational drift velocity (3) may be useful to find out if there is antigravity or not. Yet, if there is only a tiny violation of the WEP, they are presumably not sensitive enough (it would be difficult to achieve accuracy more than a few parts in $10^3$). So, in the present Appendix we suggest a method (once again we give just a rough outline) that may lead to significantly higher accuracy of one part in $10^5$.

As a first step let us find the form of a curve, such that a particle moving along the curve because of gravity has a frequency of oscillation independent of the amplitude. For frequency to be independent of the amplitude the potential energy must be $U = ks^2/2$, where the length of the arc $s$ is measured from the point of equilibrium and $k$ is a constant. The kinetic energy is then $T = m_I \dot{s}^2/2$ ($m_I$ being the inertial mass of a particle) and the frequency $\omega = \sqrt{k/m_I}$ independent of the initial value of $s$. In the gravitational field, $U = m_I a y$ where $y$ is the vertical coordinate and $a$ is the gravitational acceleration. So, we have $m_I a y = ks^2/2$ i.e.

$$y = \frac{\omega^2}{2a} s^2 \qquad (1a)$$

Using Eq. (1a) and the relation $ds^2 = dx^2 + dy^2$

$$x = \int \sqrt{\left(\frac{ds}{dy}\right)^2 - 1}\, dy = \int \sqrt{\frac{a}{2\omega^2 y} - 1}\, dy \qquad (2a)$$

Integration of (2a) is possible with a change of variable

$$y = \frac{a}{4\omega^2}(1 - \cos\alpha) \qquad (3a)$$

which together with (2a) leads to

$$x = \frac{a}{4\omega^2}(\alpha + \sin\alpha) \qquad (4a)$$

Equations (3) and (4) are parametric equations of a cycloid where the radius of the corresponding circle by which the cycloid is generated is $R = a/4\omega^2$, i.e.

$$a = 4R\omega^2 \qquad (5a)$$

Thus we have re-derived a result known from textbooks of classical mechanics that for a particle oscillating along a given cycloid, the frequency depends only on the gravitational acceleration and not on the amplitude (*i.e.* energy).

Now let us suppose that there is a tube in the form of a cycloid (Fig. 1.a) with a guiding magnetic field inside it with magnetic lines also in the form of a cycloid (one of them *AOB* is presented in Fig. 1.a). In such a tube a charged particle with sufficiently low energy ($E \leq m_I a R$, for a tube as in Fig.1.a) is trapped by gravity and presents a cycloidal pendulum (of course the influence of stray electric fields on the particle must be small compared with gravitation and effects of an inhomogeneous magnetic field must be considered).

I shall suggest two possibilities for using a cycloidal pendulum to measure the gravitational acceleration of antiprotons (and other long-lived charged particles). The first one is as follows: Let us have a particle which is oscillating along a part of cycloid *AOB* and which does not have enough energy to hit one of the detectors at points A and *B*. So, in order to detect a particle in *A* or *B* we must change its amplitude (i.e. energy) of oscillations. It may be done by resonance. Thus, detection of a particle at A or *B* is a sign that the proper frequency of the cycloidal pendulum is nearly the same as the known frequency used to obtain the effect of resonance. When we know the frequency and characteristic *R* of the cycloid, the gravitational acceleration can be calculated by Eq. (5).

Fig 1a: Antiproton forced to move as a cycloidal pendulum

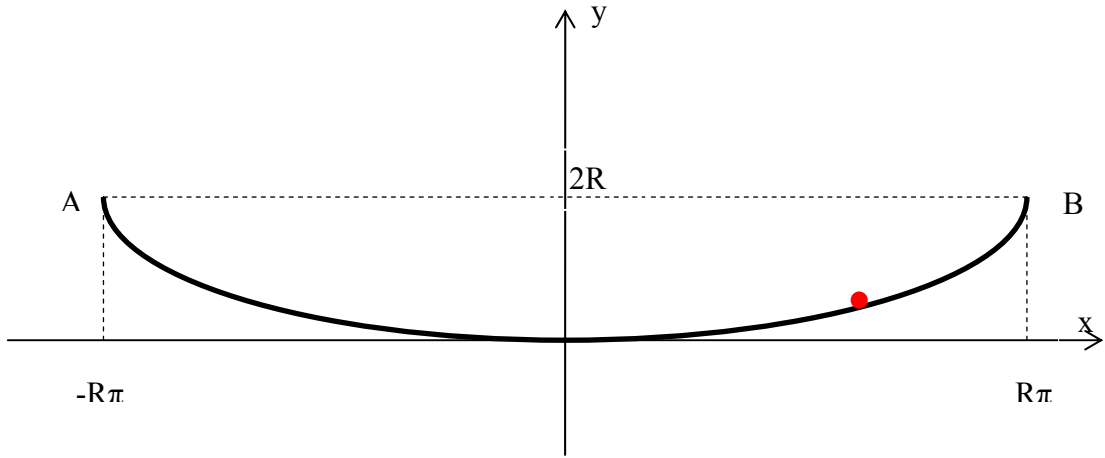

The full black line presents a drift tube in the form of cycloid with appropriate magnetic field for keeping the antiproton (red dot) in the vicinity of the axis, but allowing the particle to escape from the tube. In this way, the antiproton oscillates along the cycloid only because of gravitation. As explained in the text body, measurement of the frequency or half a period leads to determination of gravitational acceleration $a$.

The other possibility is to launch particles from O (Fig. 1.a). *A* particle with energy $E \leq m_I aR$ will return to *0* after a time $\tau$ (half of a period T) which is the same for every launched particle, because of the properties of the cycloidal pendulum. This time of flight $\tau$ would be measured and then $\omega = 2\pi/T = \pi/\tau$ i.e.

$$a = 4R\frac{\pi^2}{\tau^2} \qquad (6a)$$

In fact it is similar to the time of flight technique proposed in the PS-200 experiment with a vertical drift tube. The principal advantage of the cycloidal pendulum is that the time of flight is the same for a set of particles with different velocities, whereas in a vertical tube it depends on the initial velocity which is imprecisely known. In addition, in the variant with a cycloid the measurement can in principle be done with only one antiproton with sufficiently small energy, whereas with a vertical tube it is a statistical measurement and a large number of antiprotons are necessary.

Detailed study is under way.

## Acknowledgments

I wish to thank Alessandro Boggian from Chiasso, for his help during the preparation of this paper.